\documentstyle[prl,floats,aps,twocolumn,multicol,epsf]{revtex}
\tighten
\begin{document}
\twocolumn[\hsize\textwidth\columnwidth\hsize\csname @twocolumnfalse\endcsname
\title{Singularity deep inside the spherical charged black hole 
core} \author{Lior M. Burko} 
\address{
Department of Physics, 
Technion---Israel Institute of Technology, 32000 Haifa, Israel}
\date{\today}

\maketitle

\begin{abstract}

We study analytically the spacelike singularity inside a 
spherically-symmetric,  
charged black hole coupled to a self-gravitating spherical massless scalar 
field.   We assume spatial homogeneity, 
and find a generic solution in terms of a formal series expansion.  
This solution is tested against fully-nonlinear and 
inhomogeneous numerical simulations. We find full compliance between 
our analytical solution and the pointwise behavior of the singularity in 
the numerical simulations. This is a strong scalar-curvature 
monotonic spacelike singularity, which connects to a weak null 
singularity at asymptotically-late advanced time.  
\newline
\newline
PACS number(s): 04.70.Bw, 04.20.Dw
\end{abstract}

\vspace{3ex}
]

\section{Introduction}\label{s-int}

The spacetime singularities, which classical General 
Relativity (GR) predicts under very 
plausible assumptions to inevitably occur inside black holes 
\cite{hawking73}, are a major 
challenge for Physics, as the currently known physical laws are 
presumably invalid at such singularities. Instead, some other theory, as 
yet unknown, is expected to take over and control their structure. 
However, the GR predictions are still of the greatest importance, as they 
reveal the spacetime structure under extreme conditions in the 
strong-field regime. 

In the last few years, much advance in the study of 
black hole singularities has been achieved in the framework of classical GR. 
Poisson and Israel showed that when non-linear perturbations were allowed, 
the inner horizon of 
Reissner-Nordstr\"{o}m evolved into a scalar curvature singularity, where  
the internal mass function diverged and curvature blew up 
\cite{poisson-israel}.  
For spinning or charged black holes, it has been found that at least for 
a portion of the singularity the latter is null and weak \cite{ori91,ori92}. 
Namely, despite the divergence of spacetime 
curvature at the singularity, its tidal influence on extended physical 
objects is bounded.  The detailed structure of the null weak 
singularity was studied numerically \cite{brady95,burko97} and 
analytically \cite{bonanno95,burko-ori98} in the 
context of a spherical charged black hole, non-linearly perturbed by a 
self-gravitating, minimally-coupled, massless scalar field. (In the case 
of a spinning black hole, the evidence for the occurrence of the null 
weak singularity emerged from analytical perturbative \cite{ori92} and 
non-perturbative \cite{brady98} analyses.) It 
has been found 
numerically \cite{brady95,burko97}, that under the impact of the non-linear 
scalar field, the 
generators of the Cauchy horizon are monotonically focused, until the  
focosing effect is  
complete, and the singularity becomes spacelike, rather than null. 

Although the existence of a spacelike singularity was suggested by 
several independent 
numerical simulations \cite{gnedin93,brady95,burko97}, that 
spacelike singularity has not been shown 
analytically to exist: its very occurrence is a non-linear effect, and 
no fully non-linear analytical analyses of the problem have been 
performed. In addition, the properties of that singularity 
(both geometrical and physical) have not been studied in detail. In fact, 
the occurrence suggested by the numerical simulations is the only thing 
currently known about the spacelike singularity inside charged black 
holes. It is 
the purpose of this paper to study the spacelike singularity in the 
model of a spherical charged black hole and a neutral massless 
scalar field (the same model for which the singularity was found numerically). 

The organization of the paper is as follows. 
In Section \ref{s-model} we show that within a simplified homogeneous model 
one indeed finds a generic solution 
describing a spacelike singularity. This model is used to analyze the 
behavior of geometry and the scalar field near the singularity. 
In Section \ref{s-num} we describe fully-nonlinear and inhomogeneous 
numerical simulations which test the validity of the homogeneity 
assumption of our analytical model. 
We find full compliance between the analytical results and 
the pointwise behavior found in the numerical simulations. We also use 
the numerical simulations to study the spatial structure of the singularity. 
We conclude and discuss our results in Section \ref{s-conc}. 

\section{Analytical homogeneous model}\label{s-model}

In order to facilitate the analysis, we make the simplifying
assumption, that the singularity can be modeled to be
homogeneous. (This assumption is justified {\it a posteriori}
numerically.) That is, we assume that spatial 
gradients of the
dynamical fields are negligible near the spacelike singularity compared
with temporal gradients. Consequently, our analysis can actually
probe just the pointwise behavior at the singularity. The spatial
dependence, however, cannot be studied with this model. We shall
therefore study the spatial dependence numerically below in Sec. 
\ref{s-num}. 

We write the general homogeneous spherically-symmetric line element as 
\begin{equation}
\,ds^2=h(r)\,dt^2 +f(r)\,dr^2 +r^2\,d\Omega ^2
\label{metric}
\end{equation}
where $\,d\Omega ^2=\,d\theta ^2+\sin ^2\theta\,d\phi ^2$ is the metric on 
the unit two-sphere. Here, $r$ is the radial coordinate, defined such 
that spheres of radius $r$ have surface area $4\pi r^2$ (note 
that $r$ is timelike inside the black hole), and $t$ is normal to $r$  
($t$ is spacelike inside the black hole. There is a gauge freedom in 
$t$---see below).   
The Einstein-Maxwell-scalar equations (with a free electric field 
corresponding to charge $q$ and with a scalar field $\Phi$) are then 
given by \begin{equation}
\left(f'r+f^2-f\right)/\left(fr^2\right)=\Phi '^2  
+f\, q^2/r^4
\label{ei1}
\end{equation}
\begin{equation}
\left(h'r-hf+h\right)/\left(hr^2\right)=\Phi '^2-fq^2/r^4
\label{ei2}
\end{equation}
\begin{eqnarray}
\left[\left(h^2\right)'\,\left(f-\frac{1}{2}rf'\right)+2rf\,\left(hh''
-\frac{1}{2}h'^2\right)-2f'h^2\right]\nonumber \\ 
/\left(4rfh^2\right)=fq^2/r^4-\Phi '^2
\label{ei3}
\end{eqnarray}
in addition to the Klein-Gordon equation $\nabla^{\mu}\nabla_{\mu}\;\Phi=0$, 
whose first integral is 
$\Phi '(r)=d\;\sqrt{|f/\left(hr^4\right)|}$, where $d\ne 0$ is an 
integration constant. 
Here, a prime denotes differentiation with respect to $r$, and 
$\nabla_{\mu}$ 
denotes covariant differentiation. We seek a generic solution to 
these equations describing a spacelike singularity at $r=0$.

The source terms for the field equations [the right-hand side (RHS) of Eqs. 
(\ref{ei1})--(\ref{ei3})] contain contributions from both the electric 
field and the scalar field. We find that for a dominant electric-field 
contribution near the singularity at $r=0$, the latter is timelike, rather 
than 
spacelike. As in this case the scalar field's contribution to the 
field equations is negligible compared with the electric field's 
contribution near $r=0$, to the leading order in $r$ one would expect to 
find the same solution as with no scalar field at all. However, from the 
generalized Birkhoff theorem the solution in the latter case is nothing but 
the 
Reissner-Nordstr\"{o}m solution. Namely, with dominant electric field, to 
the leading order in $r$ the singularity is indistinguishable from the 
Reissner-Nordstr\"{o}m singularity, which is timelike. We note in passing 
that this case might be of some relevance for consideration of the 
self-gravitating scalar field for a hypothetical extension of the 
spacetime manifold {\it beyond} the weakly singular Cauchy horizon. 
A second case is scalar field and electric field which are comparable in 
strength near $r=0$. It turns out that there is no 
consistent solution for comparable 
electric- and scalar-field contributions at the singularity\footnotemark 
\footnotetext{We remark 
that this does not mean that there cannot be a solution with comparable 
contributions of both fields. We do find, however, that there is no {\it 
homogeneous} singularity with comparable contributions.}.  
The 
third case is the case where the electric field's contribution to 
the energy-momentum tensor near the singularity is negligible compared with 
the scalar field's contribution, and to this case we shall specialize 
below. We emphasize that the causal structure of the singularity at $r=0$ 
depends on which field dominates: if the electric field dominates the 
singularity is timelike, and if the scalar field dominates the 
singularity is spacelike. Note, that in this sense the spacelike 
singularity we are studying is similar to the spacelike singularity with 
vanishing charge. In fact, we find that to the leading order in $r$ the 
singularity in our case is indistinguishable from the singularity in the 
uncharged case. However, in our case the spacelike singularity is a 
non-linear effect, whereas in the uncharged case a spacelike singularity 
exists even with a vanishing scalar field (the Schwarzschild singularity). 
(There are additional differences between the spacelike singularities in 
the two cases  \cite{burko98}.) Another comment to be made about the 
timelike singularity is that it is 
reasonable to expect a {\it linear} perturbation analysis to be valid
near it, as the non-linear effects of the scalar field are negligible.

We assume a formal series expansion for the metric functions of the general 
form $f^{(n)}(r)=\sum_{i=1}^{n}f_{i}(r)$ and 
$h^{(n)}(r)=\sum_{i=1}^{n}h_{i}(r)$, and assume that both series have a 
finite radius of absolute convergence to $f$ and $h$, respectively, in the 
limit $n\to\infty$. We also assume that for all values of $i$,  
$f_i(r)=f_{i0}\,r^{m_i}$ and $h_i(r)=h_{i0}\,r^{n_i}$, with constant 
$f_{i0}$ and $h_{i0}$. (These assumptions are justified {\it a 
posteriori} numerically.) 
We then find the leading terms in $r$ for the metric 
functions, namely
\begin{eqnarray}
f(r)&=&-(\beta+1)C\,r^{\beta+2}
\nonumber \\
&+&
\frac{(\beta+1)^2(3\beta+2)}{\beta ^2}\;C^2\;q^2\;r^{2\beta+2}
+O(r^{2\beta +2 +\gamma})
\label{f}
\end{eqnarray}
\begin{equation}
h(r)=d^2C\;r^{\beta}-d^2\frac{(\beta+1)(\beta+2)}{\beta ^2}\;C^2 
\;q^2\;r^{2\beta}+O(r^{2\beta +\gamma})
\label{h}
\end{equation}
where $C$ and $\beta$ are positive parameters. Here,  
$\gamma=2$ if $\beta >2$ and 
$\gamma=\beta$ otherwise. For the scalar field we find 
\begin{eqnarray}
\Phi (r)&=&\sqrt{\beta +1}\;\ln r-\frac{(\beta +1)^{3/2}(3\beta+2)}
{2\beta^3}\; C^2\; q^2\; r^{\beta}
\nonumber \\
&+&O(r^{\beta +\gamma}) .
\label{Phi}
\end{eqnarray}
This form for the solution suggests that $f,h$ are given by infinite 
double series 
\begin{equation}
f,h=\sum_{m=1}^{\infty}\sum_{n=1}^{\infty}a_{mn}^{f,h}\; r^{\beta 
m+(\beta +2)(n-1)+\sigma _{f,h}} ,
\label{series}
\end{equation}
$a_{mn}^{f}=a_{mn}^{f}(C,\beta )$ and 
$a_{mn}^{h}=d^2\tilde{a}_{mn}^{h}(C,\beta )$ being the expansion 
coefficients. Here, $\sigma _{f}=0$ and $\sigma _{h}=-2$. \footnotemark 
\footnotetext{We note 
that whereas with vanishing electric charge it has been shown that the 
expansion coefficients in the series expansion analogous to 
(\ref{series}) can be found uniquely for all orders \cite{burko-proc}, 
in the case studied 
here this is yet to be proved. However, we believe that this is also the 
case here, despite the formal difference in the series expansions of the 
two cases: whereas in the uncharged case the expansion is given in the 
form of a simple series, in the charged case it is given in terms of a 
double series.} As we have already remarked, the black hole electric charge 
does not appear in the leading order terms of either the metric functions 
or the scalar field. This is a direct consequence of the dominance of the 
scalar field over the electric field near the singularity. Consequently, 
the derivation of the solution is similar to the derivation in 
\cite{burko98}, and further details can be found there. The charge 
affects the spacetime geometry and the behavior of the scalar field in 
higher order terms, however, as is evident from the solution, such that 
at any finite value of $r$, even very close to the singularity, both the 
geometry and the field are different from the uncharged case's 
counterparts. Note, that we find the parameter $\beta$ to be positive. In 
the case of the spacelike singularity inside an uncharged black hole, the 
corresponding parameter is found to be greater than $-1$, which is the 
value found for the Schwarzschild singularity. Thus, in the uncharged 
case the parameter $\beta$ changes continuously from its vacuum value, 
whereas in the charged case we study here it jumps discontinuously from 
the $-2$ value of the Reissner-Nordstr\"{o}m singularity which 
corresponds to the (electro-)vacuum limit of our spacetime, to positive 
values. This discontinuity is the result of the non-linear nature of the 
spacelike singularity: with no scalar field (the electro-vacuum case) the 
singularity at $r=0$ is timelike. With the addition of a scalar field the 
only singularity we find is spacelike. The change from timelike to 
spacelike is discontinuous, and this is what we indeed find.    

\begin{figure}
\input{epsf}
\centerline{\epsfysize 7.0cm
\epsfbox{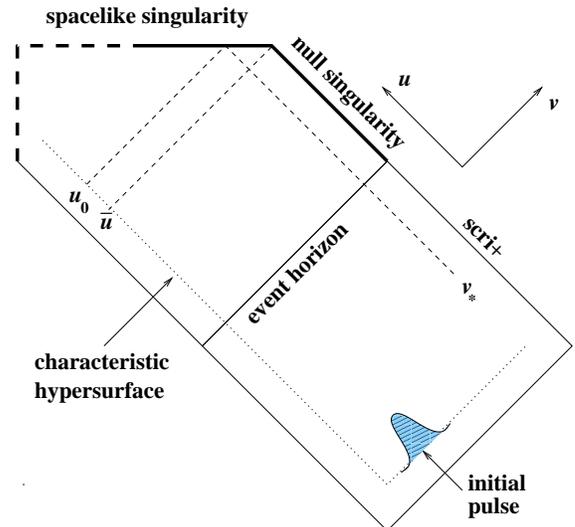}}
\caption{The Penrose diagram of the simulated spacetime. 
Singularities are depicted by thick lines. Dashed thick lines denote 
regions of the singularity not covered by the domain of integration,
and describe a possible picture of these portions. The causal structure 
of the singularity changes from null to spacelike at 
$(u=\bar{u},v=\infty )$. For any outgoing null ray at $u=u_0 > \bar{u}$ 
the spacelike singularity is reached at $v=v_*$.} 
\label{fig0}  
\end{figure}

From this solution, we draw the following inferences: First, we define a 
`tortoise' 
coordinate $r_*$ by $g_{r_*r_*}=-g_{tt}$. Then, null coordinates are 
defined by $t=(v-u)/2$ and $r_*=(v+u)/2$, with ingoing $v$ and outgoing 
$u$. For a certain outgoing ray $u=u_0\equiv \rm{const}$ which runs into the 
spacelike singularity, the latter is hit at some finite value of 
advanced time $v=v_*$ (see Fig. \ref{fig0}). (The null singularity 
is located at $v=\infty$.) 
We find that to the leading order in $v_*-v$  
\begin{equation} 
r(u_0,v)=\left[d^2/(\beta +1)\right]^{1/4}\;(v_*-v)^{1/2}.
\label{r(v)}
\end{equation}
This property of the spacelike singularity inside a non-linearly 
perturbed charged black hole is similar to the properties of the 
singularity inside uncharged black holes \cite{burko98}. 
We next consider an observer who follows a radial $t={\rm const}$ 
trajectory. Let $\tau$ be that observer's proper time, set such that 
$\tau =0$ at $r=0$. We denote by $x^0$ the coordinate tangent to the 
worldline. Then, to the leading order in $\tau$, the $(00)$ 
tetrad component of the Ricci tensor and the Kretschmann scalar, 
correspondingly,  are given by 
\begin{equation}
R_{(0)(0)}(\tau)=8\;\frac{\beta +1}{(\beta+4)^2}\;\frac{1}{\tau ^2}
\label{ricci}
\end{equation}
\begin{equation}
R_{\alpha\beta\gamma\delta}R^{\alpha\beta\gamma\delta}(\tau )=
64\; \frac{2\beta (\beta +1)+3}{(\beta +4)^4}\;\frac{1}{\tau ^4} .
\label{kre}
\end{equation}
As Eqs. (\ref{ricci}) and (\ref{kre}) are given to the leading order in 
$\tau$, they are identical to the corresponding equations in the uncharged 
case (cf. \cite{burko98}). 
Equation (\ref{kre}) implies that this is a scalar-curvature singularity. 
From Eq. (\ref{ricci}) and from the theorem by Clarke and Kr\'{o}lak 
\cite{clarke86} it 
then follows that this singularity is strong in the Tipler sense 
\cite{tipler77}, as the twice integrated $R_{(0)(0)}(\tau)$ over $\tau$ 
diverges logarithmically as $\tau \to 0$. Namely, 
any extended physical object will unavoidably be crushed to zero volume 
upon arrival to the singularity.  As we find that $\beta >0$, we infer  
that there is no finite $\beta$ which nullifies any of the RHS's of 
either Eq. 
(\ref{ricci}) or Eq. (\ref{kre}). Consequently, for the entire range of 
permissible parameters, this is a strong scalar-curvature singularity. 
This property of the singularity is in sharp contrast with the Cauchy 
horizon singularity, which is known to be null and weak.  

Let us consider the genericity of our solution. The notion of a general 
solution for non-linear differential equations is not unambiguous. 
However, we can count the number of arbitrary parameters in our solution, 
and thus show that it is generic. From the physical point of view, one 
would expect two arbitrary parameters (one because of 
Birkhoff's theorem and one due to the scalar field. 
Note that as the scalar field is neutral, the 
electric charge is fixed).  Apparently, our solution has three 
arbitrary parameter, namely $\beta$, $d^2$, and $C$. However, the 
arbitrariness in $d^2$ reflects merely a trivial gauge mode, i.e., the 
freedom to re-scale the coordinate $t$ (see Ref. \cite{burko98}).  
Consequently, our solution has 
two physical degrees of freedom, and is therefore generic. These 
considerations do not rule out the possible existence of other generic 
solutions. However, this solution is the one which is realized 
in the numerical simulations (see below).

\section{Fully inhomogeneous numerical simulations}\label{s-num}
 
The analytical approach presented above assumes the singularity to be 
homogeneous. However, the spacetime inside black holes (even very close to 
the 
singularity) is expected in general to be inhomogeneous. To what 
extent then is the homogeneity assumption restrictive? 
We shall present in what 
follows the results of fully non-linear and inhomogeneous numerical  
simulations. We show, that the above homogeneous analytical model 
succeeds to describe the {\it pointwise} behavior of the 
singularity as inferred from the inhomogeneous 
simulations. In this sense the assumption of homogeneity is a 
simplification which does not ruin the structure of the singularity, as 
the numerical simulations reveal that the dependence of the geometry at 
and near the singularity depends only weakly on spatial coordinates: it is 
the temporal gradients which dominate. However, the spatial dependence 
can still be studied numerically. We shall do just this after we confront 
the predictions of the analytical model with the pointwise behavior of 
the numerical simulations.

Let us discuss then the results obtained from fully non-linear (and 
inhomogeneous) numerical 
simulations. We used the same code which was used in Ref. \cite{burko97} 
(see also \cite{burko-ori97}) to study the null singularity in this model. 
This code is based on double-null coordinates and on free evolution 
of the metric functions and fields. The code is stable and converges 
with second order. (Slight adaptations of the code were needed, to allow 
for a careful approach to $r=0$. These modifications are described in 
Ref. \cite{burko98}.) Our initial value set-up is such that prior to the 
smooth initial pulse of compact support the geometry is 
Reissner-Nordstr\"{o}m with charge $q$ and mass $M$ (see Fig. 
\ref{fig0}). Then, due to the 
scalar field the geometry is changed, and the final mass of the black 
hole is $M_f$. 

\begin{figure}
\input{epsf}
\centerline{\epsfysize 7.0cm
\epsfbox{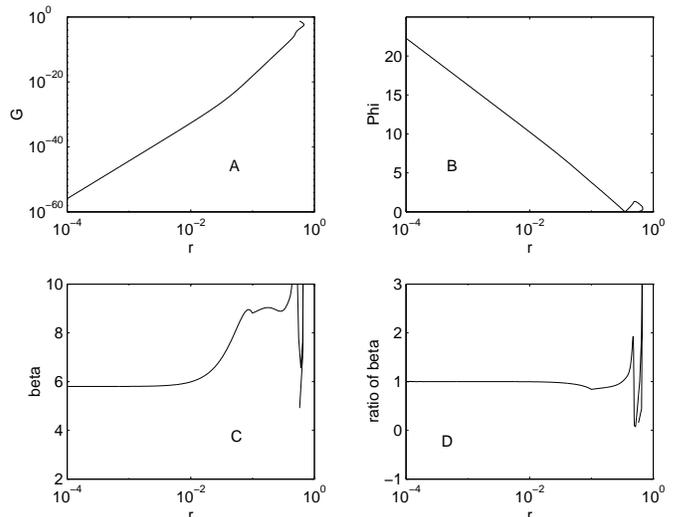}}
\caption{(A) The metric function $G=-2g_{uv}$. (B) The scalar field 
$\Phi$. (C) The parameter $\beta$ as calculated from the exponent of $h$. 
(D) The ratio of $\beta$ as calculated from the exponent of $h$ and 
$\beta$ as calculated from 
the amplitude of $\Phi$. In all graphs the abscissae are $r$.} 
\label{fig1} 
\end{figure}

\begin{figure}
\input{epsf}
\centerline{\epsfysize 7.0cm
\epsfbox{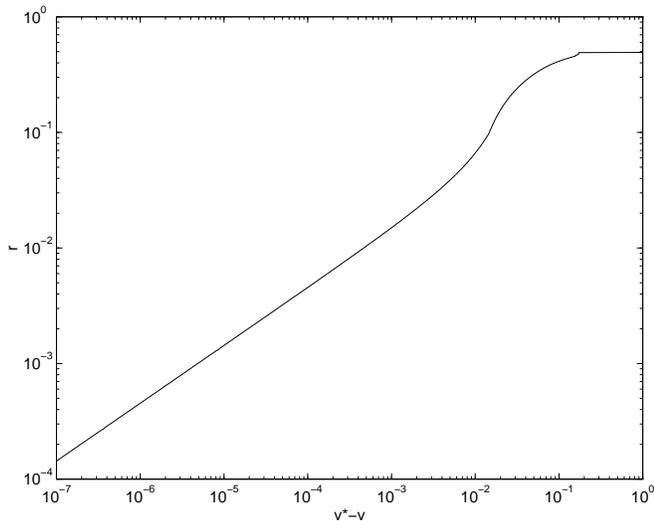}}
\caption{The radial coordinate $r(u_0,v)$ as function of (advanced-time) 
distance $v_*-v$ from the
spacelike singularity. The singularity is located at $v=v_*$.}
\label{fig2}
\end{figure}

\begin{figure}
\input{epsf}  
\centerline{\epsfysize 7.0cm
\epsfbox{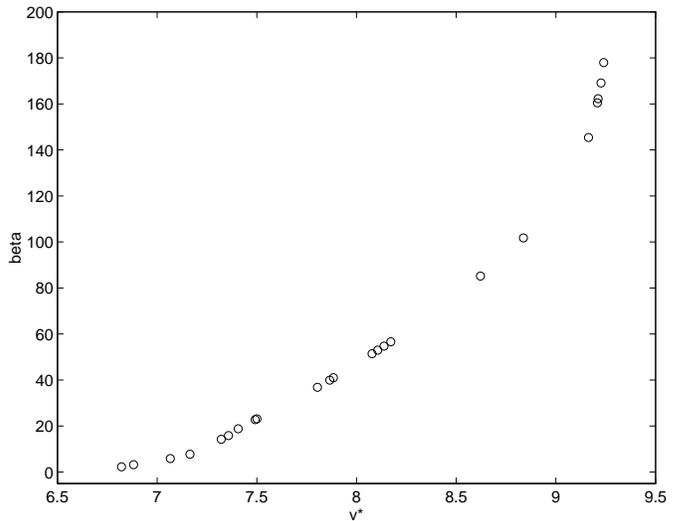}}
\caption{The value of $\beta$ for different outgoing null rays as a
function of $v_{*}$, namely, as a function of the value of the ingoing
coordinate $v$ for which the spacelike singularity is reached when
approached along outgoing null rays.}
\label{fig3}
\end{figure}

Figures \ref{fig1}, \ref{fig2} and \ref{fig3}  
display the results of the numerical simulations with 
$q/M=0.998$ and
$M_f/M\approx 1.51$. We have also performed numerical simulations 
with other values of the parameters, and found no qualitative change 
of the behavior. In fact, when we increase the amplitude of the 
scalar-field pulse (while keeping the support of the pulse fixed), we find 
the following two effects. First, the 
ratio $M_f/M$ increases, as should be expected from the greater energy 
content of the pulse. Second, the larger the amplitude, the earlier the 
spacelike singularity occurs. Namely, with larger amplitude for the 
initial pulse the generators of the Cauchy horizon focus faster, and the 
Cauchy horizon focuses completely within a smaller lapse of affine 
parameter. However, the spacelike singularity itself is unchanged by this 
change of the initial parameters, and is in this sense robust. Very 
similarly, varying the value of $q/M$ does not change the properties of 
the singularity. This should indeed be expected from our analytical 
considerations of Section \ref{s-model}: the singularity is dominated by 
the scalar field, and to the leading order in $r$ the charge of the black 
hole has no effect.  

The data are shown along an outgoing ray at 
$u_0$  
approaching the spacelike singularity at $r=0$ (similar results are 
obtained for any choice of such an outgoing ray; in all four graphs of 
Fig. \ref{fig1} the abscissae are $r$): \ref{fig1}(A) the 
metric component $g_{uv}$, for double-null coordinates defined such that 
$v=r$ on the outgoing segment of the characteristic hypersurface, and 
$u\propto -r$ on its ingoing segment. 
The vanishing of $g_{uv}$ implies the vanishing of both 
$f$ and $h$. Moreover, the asymptotically linear 
curve in this plot implies a power-law behavior near the singularity. 
(This is checked more accurately in Fig. 
\ref{fig1}(C).) Figure \ref{fig1}(B) displays the scalar field $\Phi$. We 
find that $\Phi$ diverges logarithmically, in agreement with Eq. 
(\ref{Phi}). Figure \ref{fig1}(C) 
shows the local power index of $g_{uv}$, namely, the local slope in Fig. 
\ref{fig1}(A). The asymptotically constant value to which the local power 
approaches---in agreement with Eqs. (\ref{f}) and (\ref{h})---implies that 
the power-law assumption for the metric functions 
is justified. In fact, this power-law index is just the value of the 
parameter $\beta$ for the intersection point of the outgoing 
ray under consideration with the spacelike singularity. Finally, we test  
another aspect of the analytical model. We found that the same parameter 
$\beta$ appears in two different contexts, namely, in the exponent of the 
metric functions [Eqs. (\ref{f}) and (\ref{h})], and in the amplitude of the 
scalar field [Eq. (\ref{Phi})]. The ratio of 
these two values for $\beta$ is predicted, therefore, to asymptotically 
approach a value of unity. This is indeed shown in Fig. \ref{fig1}(D), to 
within $1$ part in $10^{3}$.  

Figure \ref{fig2} shows $r(u_0,v)$ as a function of the  
distance  (in terms of advanced time) from the singularity $v_*-v$. 
The slope is equal 
asymptotically to the value of $1/2$ implied by Eq. (\ref{r(v)}) 
to within a numerical error of $1\%$.

Finally, we use the numerical simulation in order to check the spatial 
dependence of the solution. Figure \ref{fig3} displays the value of the 
parameter $\beta$ as a function of advanced time, namely, as a function 
of the value of $v_*$ at which the outgoing ray under 
consideration hits the singularity. We find that along the spacelike 
singularity, upon approach 
to the spacetime event where the causal structure of the singularity changes 
from null to spacelike, $\beta$ increases rapidly. This suggests that 
at the particular outgoing ray at $u=\bar{u}$ 
for which that event is reached $\beta$ may be  
infinite. We can therefore compare the relative strength of 
different points along the spacelike singularity. Although this 
singularity is strong throughout, the nearer one is to the spacetime 
event where the causal structure changes, the weaker is the strong 
singularity, as curvature scalars diverge slowlier. 
That is, let us consider a constant $r$ hypersurface near the singularity 
(namely, deep in the asymptotic region). Then, approaching $\bar{u}$ 
along this hypersurface curvature scalars decrease (as $\beta$ 
increases).  
This fits well with our overall picture: when the singularity 
first forms it is null and weak. At later times, the weak null singularity becomes 
stronger with increasing affine parameter. This is manifested by the 
focusing effect, and by the inverse proportion of the blue-shift factors 
to the area coordinate \cite{burko-ori98}. Thus, we have a weak 
singularity which becomes stronger. Then, at $u=\bar{u}$, $r$ vanishes and 
the causal structure of 
the singularity changes and it becomes spacelike and  
strong, and strengthens as one gets farther from the changing point, 
namely, as $u-\bar{u}$ increases. 

\section{Discussion}\label{s-conc}

We find analytically a simple homogeneous model, which describes the  
pointwise behavior of the 
spacelike singularity which occurs inside spherical charged black 
holes with a non-linear scalar field. This singularity is generic in the 
sense that the solution has the expected number of free parameters. The 
numerical simulations imply that the 
singularity we find in the simple homogeneous model successfully describes 
the pointwise behavior of the singularity     
of the fully nonlinear and in general inhomogeneous spacetime. The 
spacelike singularity is scalar curvature (like the null singularity 
which precedes it), monotonic, and strong (unlike the null singularity, 
which is weak).  

It remains an open question, to 
what extent this simple model captures the essence of the singularity 
inside realistic black holes, namely, spinning black holes with vacuum 
perturbations: First, scalar fields are known to be related to unique 
phenomena, such as the destruction of the oscillations of the 
Belinskii-Khalatnikov-Lifshitz (BKL) 
singularity \cite{belinskii73}. Second, we find that the scalar field in 
our model is dominant near the singularity, whereas the BKL 
singularity  is often dominated by vacuum perturbations rather than by 
matter fields \cite{landau}. 
Finally, both the spherical symmetry and the scalar field are 
mere toy models for the more realistic cases. A major open question is then 
whether in more realistic cases there would 
indeed be a spacelike singularity to the future of 
the null singularity, and whether it would be monotonic -- like the 
spacelike singularity inside spherical charged black holes with a scalar 
field -- or oscillatory -- like the BKL singularity.

\section*{Acknowledgements}

I am indebted to Amos Ori for many invaluable discussions and helpful 
comments. 
This research was supported in part by the United States--Israel 
Binational Science Foundation.

\end{document}